\documentclass[]{spie}  

\usepackage{amsmath,amsfonts,amssymb}
\usepackage{graphicx}
\usepackage{subfig}
\usepackage{upgreek}
\usepackage[colorlinks=true, allcolors=blue]{hyperref}
\graphicspath{{figures/}}
\title{SPIE Proceedings: \\
GALI a Gamma-ray Burst Localizing Instrument}

\newcommand{\sipm}{SiPM}
\newcommand{\csi}{CsI(Tl)}
\newcommand{\gagg}{Ce:GAGG}
\newcommand{\sensl}{Sensl J60035}
\newcommand{\hama}{Hamamatsu 14160-6050HS}
\newcommand{\cs}{$^{137}$Cs}

\newcommand{\am}{$^{241}$Am}

\newcommand{\sio}{SiO$_2$}

\author[a]{Roi Rahin}
\author[a]{Luca Moleri}
\author[a]{Alex Vdovin}
\author[a]{Amir Feigenboim}
\author[a]{Solomon Margolin}
\author[a]{Shlomit Tarem}
\author[a]{Ehud Behar}
\author[b]{Max Ghelman}
\author[b]{Alon Osovizky}
\affil[a]{Department of Physics, Technion, Haifa, Israel}
\affil[b]{Nuclear Research Center Negev, Israel}

\authorinfo{Further author information: (Send correspondence to R.R)\\R.R.: E-mail: roir@campus.technion.ac.il}

\begin{document} 
\maketitle

\begin{abstract}
The detection of astrophysical Gamma-Ray Bursts (GRBs) has always been intertwined with the challenge of identifying the direction of the source. Accurate angular localization of better than a degree has been achieved to date only with heavy instruments on large satellites, and a limited field of view.
The recent discovery of the association of GRBs with neutron star mergers gives new motivation for observing the entire $\gamma$-ray sky at once with high sensitivity and accurate directional capability.
We present a novel $\gamma$-ray detector concept, which utilizes the mutual occultation between many small scintillators to reconstruct the GRB direction. We built an instrument with 90 (9\,mm)$^3$ \csi~scintillator cubes attached to silicon photomultipliers. Our laboratory prototype tested with a 60\,keV source demonstrates an angular accuracy of a few degrees for $\sim$25 ph\,cm$^{-2}$ bursts. 
Simulations of realistic GRBs and background show that the achievable angular localization accuracy with a similar instrument occupying $1$l volume is $<2^\circ$. The proposed concept can be easily scaled to fit into small satellites, as well as large missions.
\end{abstract}


\section{INTRODUCTION}
\label{sec:intro}  

\subsection{Astrophysical Context}
Astrophysical $\gamma$-ray Bursts (GRBs) are the most remarkable transient phenomena in high-energy astrophysics. Long GRBs, those lasting more than 2\,s, are generally associated with supernovae, the explosion following the collapse of massive stars. Short GRBs, lasting less than 2\,s, are  associated with the coalescence of a neutron stars binary \cite{Blinnikov84,paczynski1986gamma}, and could also result from a merger of a NS and a black-hole binary \cite{mochkovitch1993gamma}. 

Following the first discoveries of gravitational waves (GWs) from compact stellar mergers by Advanced LIGO (Laser Interferometer Gravitational-wave Observatory), the astrophysics community has concentrated its efforts on detecting their electromagnetic (EM) counterparts. The first direct detection of GWs by the two LIGO facilities occurred in September 2015 \cite{abbott2016observation}.
In August 2017 the European VIRGO detector joined the LIGO observation run \cite{abbott2017gw170814}. The combined detection by three interferometers potentially improves the event localization from more than a thousand square degrees to tens of square degrees, depending of course on the strength of the signal.

The first EM counterpart of a GW event was observed on 17 August, 2017. This LIGO-VIRGO GW event \cite{abbott2017gw170817}) was followed  1.7\,s later by a short GRB (GRB\,170817A) independently detected by the Fermi $\gamma$-ray Burst Monitor (GBM) and by IBIS on board INTEGRAL \cite{abbott2017gravitational}. Many other facilities around the world started to search for the site of the event in all possible wavebands. It took approximately 11 hours before the source was identified in the lenticular galaxy NGC 4993 \cite{abbott2017multi}. 

Increasing the number of simultaneously observed GW and EM counterparts is paramount to addressing fundamental questions on the nature of coalescing neutron stars and black-holes. The goal of the method described in the present paper is to detect GRBs at high sensitivity and to identify their direction with high accuracy. Such capabilities will not only enable fast follow-up with telescopes, but will also allow LIGO-VIRGO to search for sub-threshold events once the time and direction are known.

\subsection{GRB Localization}
\label{subsec: GRB localization}
A single soft $\gamma$-ray detector unit generally cannot identify the direction of incident photons. Angular localization of GRBs thus require multiple detectors. 
The coded-mask aperture method, most successfully implemented on Swift-BAT \cite{barthelmy2005burst}, is an array of detectors partially covered by a mask. The coded mask generates different shading patterns over the detectors array, varying with the source direction. This technique enables the reconstruction of the direction of a GRB up to 20' (minutes of arc). However,
it requires a large detector area and volume due to the required separation between the mask and the array. Moreover, it has a limited field of view, and the mask blocks many of the source photons. 

Another method is a sparse array of scintillators, e.g., on Compton-BATSE \cite{Fishman85} and Fermi-GBM \cite{meegan2009fermi}, which are distributed over the spacecraft, each facing a different direction. The relative signals in each scintillator
provide information about the direction of the source. For example, those facing the opposite direction are shielded by the satellite and will have lower count rates. 
Angular localization of soft $\gamma$-ray sources can
also be achieved if the detectors are far enough apart, as on different space crafts, and the
different times of arrival can be discriminated.\cite{fiore2018hermes} Alternatively, direction reconstruction is possible if the detection is time-coincident with
another observation of a different messenger (e.g. visible light), where better angular
accuracy is available.

\section{The GALI concept}
\begin{figure}[h]
\begin{center}
\subfloat{
\includegraphics[scale=0.48]{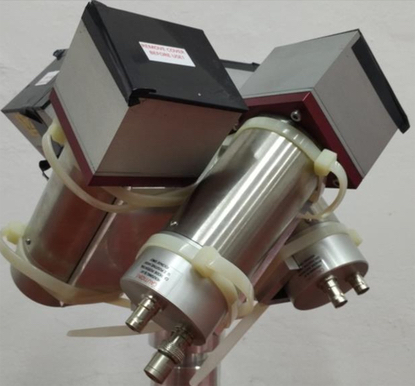}}
\subfloat{
\includegraphics[scale=0.27]{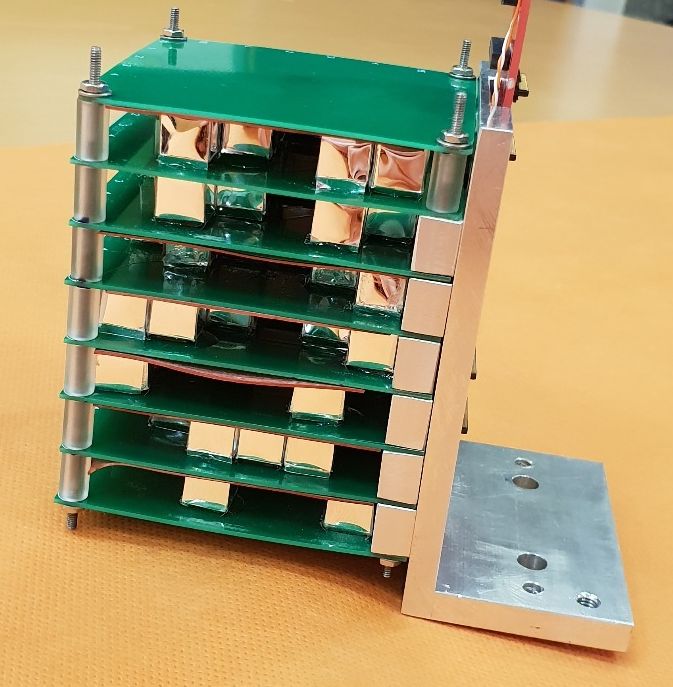}}
\end{center}
\caption {Two prototype instruments demonstrating the two conceptual detector designs \textbf{Left}: The GTM - a traditional design based on large, asymmetric box-shaped scintillators with various orientations. This design is similar to the Fermi/GBM.   \textbf{Right}: The GALI design, which uses (9\,mm)$^3$ cubic scintillators in a 3D structure that exploits their different mutual occultation from different directions.}
\label{fig:concept}
\end{figure}

In traditional astrophysical $\gamma$-ray detectors, scintillators are built with different cross-sections towards different directions to produce a gradually varying response with angle. These systems rely on the scintillators facing various orientations to reconstruct the direction of the source. 
In contrast, the $\gamma$-ray-burst Localizing Instrument (GALI) concept presented here exploits \textit{mutual occultation} of numerous small scintillators, distributed within a small volume, to provide directional information.
The method relies on the entire array looking significantly different from different directions. Due to the occultation, the count rates from each scintillator will vary dramatically as a function of the source direction, even for small angle differences.
In a sense, this is similar to the coded mask aperture method, but the mask itself is composed of detecting scintillators, so that no precious photons are lost. 
The low count rates in each individual small scintillator are compensated by the large number of scintillators.
As in traditional approaches, the sensitivity to weak sources depends on the total size of the detecting volume. 
Fig.\,\ref{fig:concept} shows two prototype instruments demonstrating the differences between a traditional detector design and the present concept.
The first is the Gamma ray Transient Monitor (GTM) whose design is similar to the Fermi/GBM, and which has been described in past SPIE proceedings \cite{yacobi2018gamma}. The second is the GALI laboratory prototype described in section \ref{subsec: 90 units detector prototype}.

The GALI laboratory prototype presented here is composed of (9\,mm)$^3$ cubic scintillators. These are read out by Si Photo-Multipliers (\sipm s), which obviates the traditional, cumbersome, Photo-Multiplier Tubes (PMTs). Adopting such compact light sensors enables filling a volume with a large number of small detectors at the expense of the size of each individual one. An additional advantage of \sipm s over PMTs is their significantly lower operation voltage (tens of volt instead of hundreds).  

A simulation of the system performance is presented in section~\ref{sec:sim}, a detailed description of the experimental setup is given in section ~\ref{sec: laboratory experiments}, and the results of the experiments are given in section \ref{sec:results}.

\section{Simulations} \label{sec:sim}
To check the performance of the GALI concept we run simulations using MEGAlib\cite{zoglauer2006megalib}, a simulation software based on Geant4\cite{agostinelli2003geant4}. In these simulations all $\gamma$-ray interactions with scintillators are considered as counts in the relevant scintillator. Each simulation run is composed of two steps: A bright GRB with no background and the background simulation.
We use the bright GRB simulation to estimate the average counts on each scintillator in the system with no background: a burst of 1000~ph$\cdot$cm$^{-2}$ , whose photon spectrum ($dN/dE$) is a band function\cite{band1993batse} (equation \ref{Band}) with $\alpha=1.1$, $\beta=2.3$ and $E_{peak} =(2+\alpha)E_0 =  266$~keV.
\begin{equation} \label{Band}
\frac{dN(E)}{dE}= A\begin{cases} {(\frac{E}{100keV})^\alpha \exp \left( { - \frac{E}{E_0}} \right)}  & E \le (\alpha - \beta) E_0\mbox{ } \\ {\left[{\left( {\alpha - \beta } \right)\frac{E_0}{100keV}} \right]^{\left( {\alpha - \beta } \right)} (\frac{E}{100keV})^\beta \exp \left( {\beta - \alpha } \right)}  & E > (\alpha - \beta) E_0\mbox{ } \end{cases}
\end{equation}
The simulations are repeated varying the angular coordinates at 5$^\circ$ intervals within the upper hemisphere. The average counts on each scintillator are then interpolated at either 0.5$^\circ$ or 1$^\circ$ intervals. We divide the average counts by the total number of counts for each burst to obtain the relative average counts on each scintillator. The result is an array of relative counts for each angle in the hemisphere.
The background simulation is based on various lower earth orbit (LEO) observations of hard X-rays and $\gamma$-rays \cite{gruber1999spectrum,mizuno2004cosmic,ajello2008cosmic,abdo2009fermi,turler2010integral} and is included in the MEGAlib software package. Background caused by Leptonic and Hadronic components was not included, being estimated in simulation to be less than 1\% of the photonic background, far less than the statistical uncertainty. The Hadronic background may need to be simulated later depending on the spacecraft platform which will host the experiment. 

After running the simulations, we estimate the directional capabilities of the detector system. We consider 1 second bursts of 10~ph cm$^{-2}$ s$^{-1}$ from a given direction in the presence of the background. The burst is generated using the aforementioned relative average counts by applying poisson statistics to the expected average of each scintillator. We then add the background with poisson uncertainty. We reconstruct the burst direction using a cstat estimator \cite{cash1979parameter} between the generated counts and the simulated ones at each angle. The reconstructed direction is that which gives minimal cstat value. The burst generation and direction reconstruction process is repeated 100 times for each angle at 5$^\circ$ intervals so as to generate an accuracy map of the entire hemisphere.

\begin{figure} [ht]
\begin{center}
\includegraphics[scale=0.25]{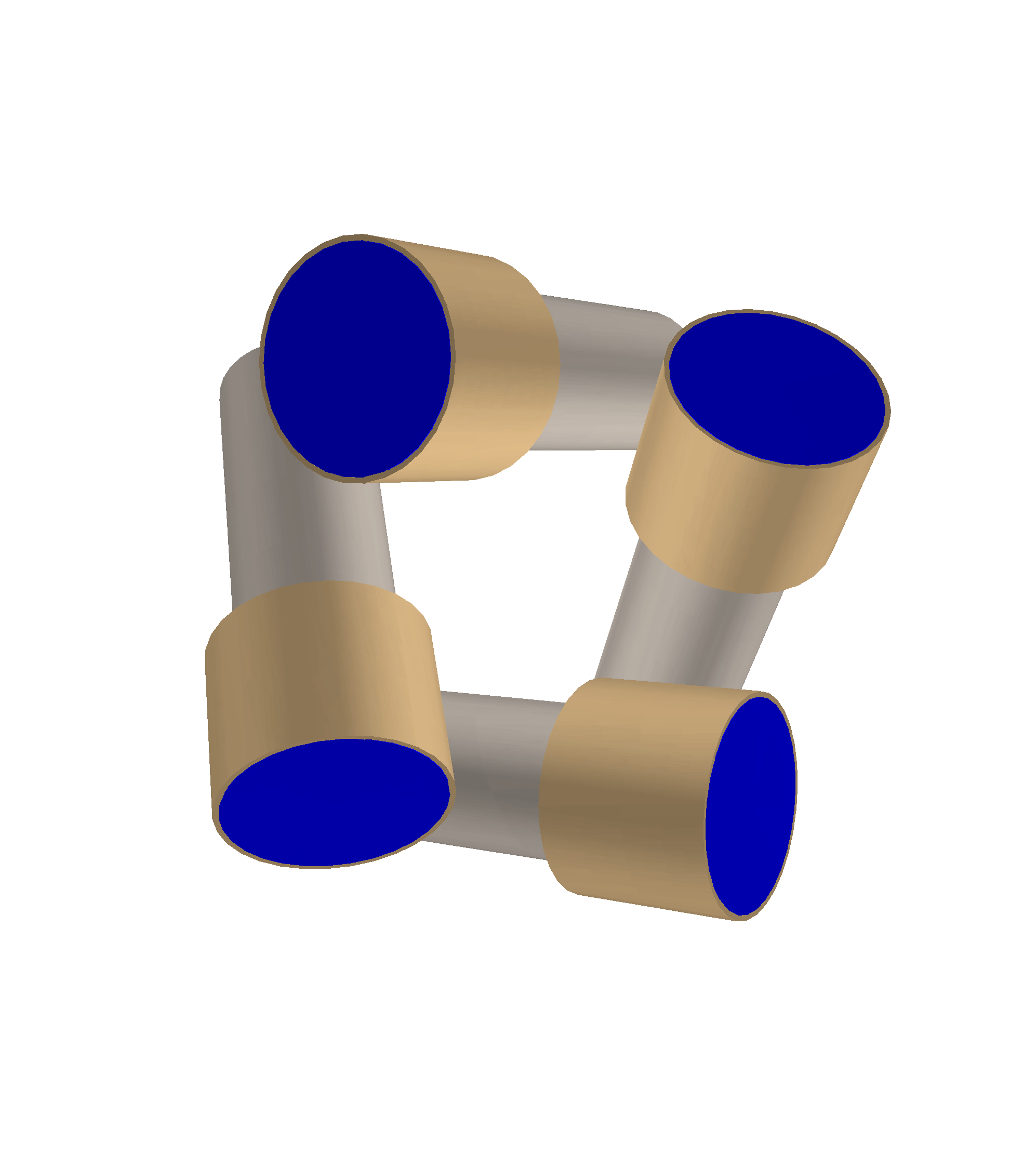}
\includegraphics[scale=0.25]{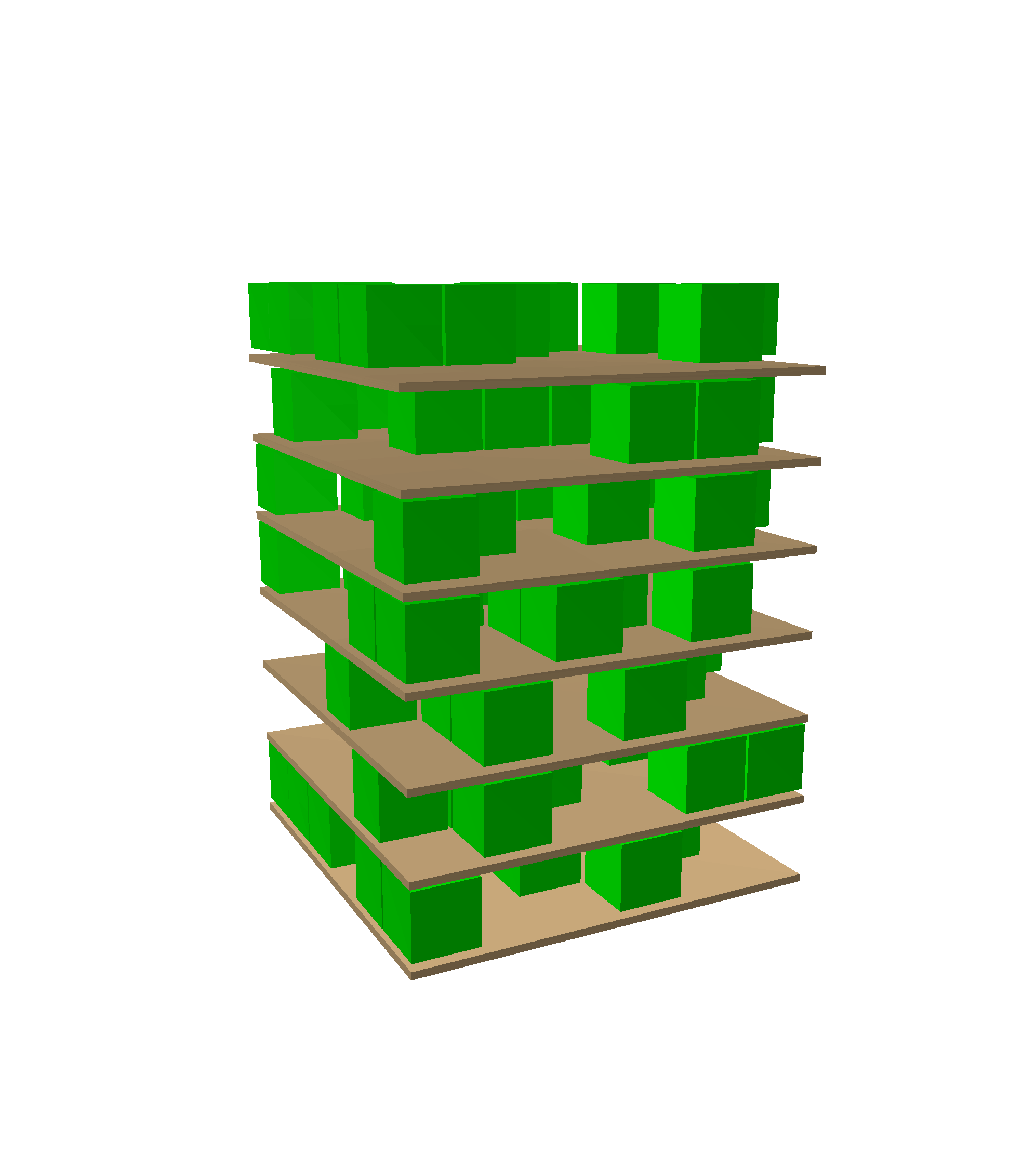}
\includegraphics[scale=0.25]{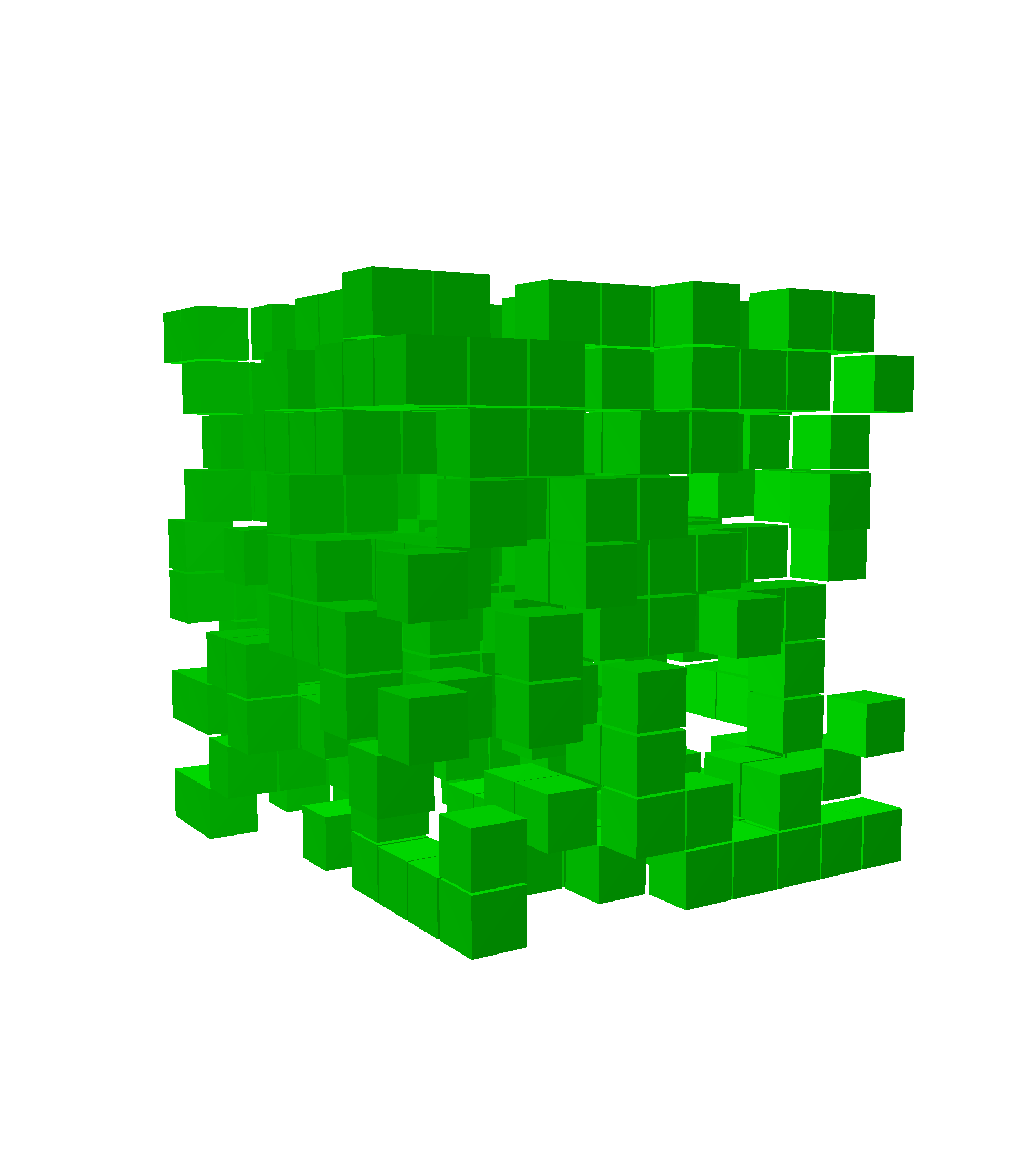}
\end{center}
\caption[example] 
{ \label{fig:det_sys} Three detector configurations compared using simulations. \textbf{Left}: A GTM configuration of four 3" diameter 1" thick cylindrical detectors - the GTM. \textbf{Middle}: A GALI configuration of ninety (9\,mm)$^3$ cubic scintillators. Between the detectors are PCB boards. This system is simulated as close as possible to a lab model. \textbf{Right}: A GALI configuration of 350 (9\,mm)$^3$ cubic scintillators.}
\end{figure}

The simulated detectors are two different GALI systems: a random configuration of 90 (9\,mm)$^3$ scintillator cubes spread in a 6$\times$6$\times$7~cm$^3$ volume and one of 350 cubes in a 10$\times$10$\times$10~cm$^3$ volume. 
To compare the potential improvement of the GALI accuracy with respect to previous existing concepts, we also simulated the a GTM detector with four 3" diameter 1" thick cylindrical scintillators. The GTM has similar effective area compared to the 350-scintillator GALI, but has approximately twice that of the 90-scintillator GALI. The simulated systems are shown in figure \ref{fig:det_sys}. Notice that the 90-scintillator GALI contains the PCB boards upon which the detectors are mounted as in the the lab prototype described in section~\ref{subsec: 90 units detector prototype}, whereas no 350-scintillator GALI prototype exists, so no PCB boards have been included in this case. 

\begin{figure} [ht]
\begin{center}
\includegraphics[scale=0.25]{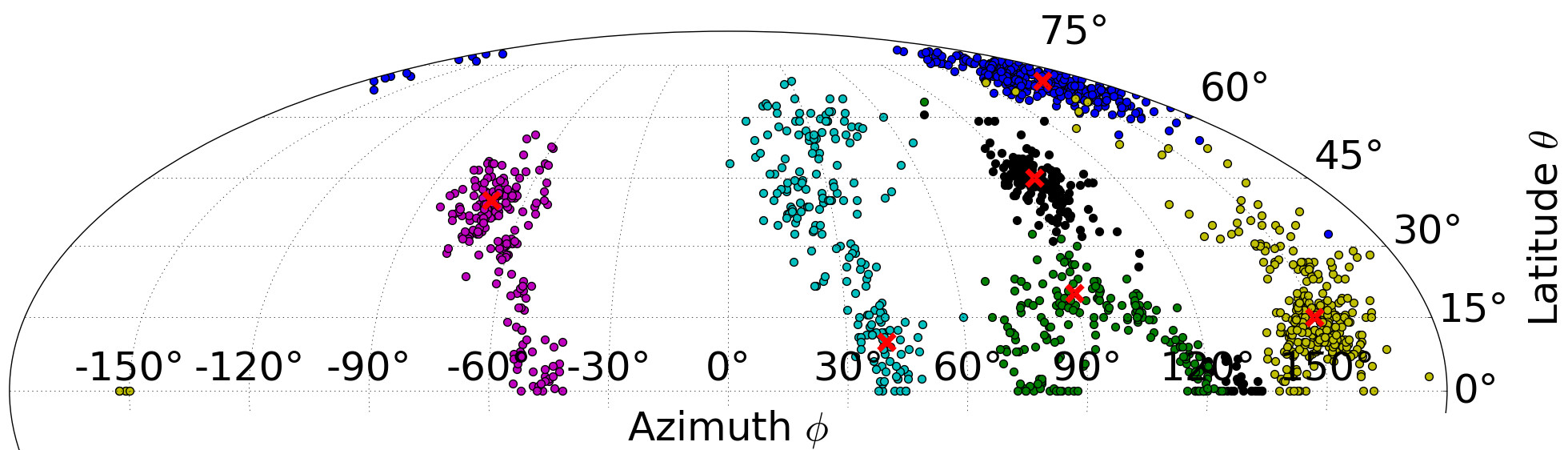}
\includegraphics[scale=0.25]{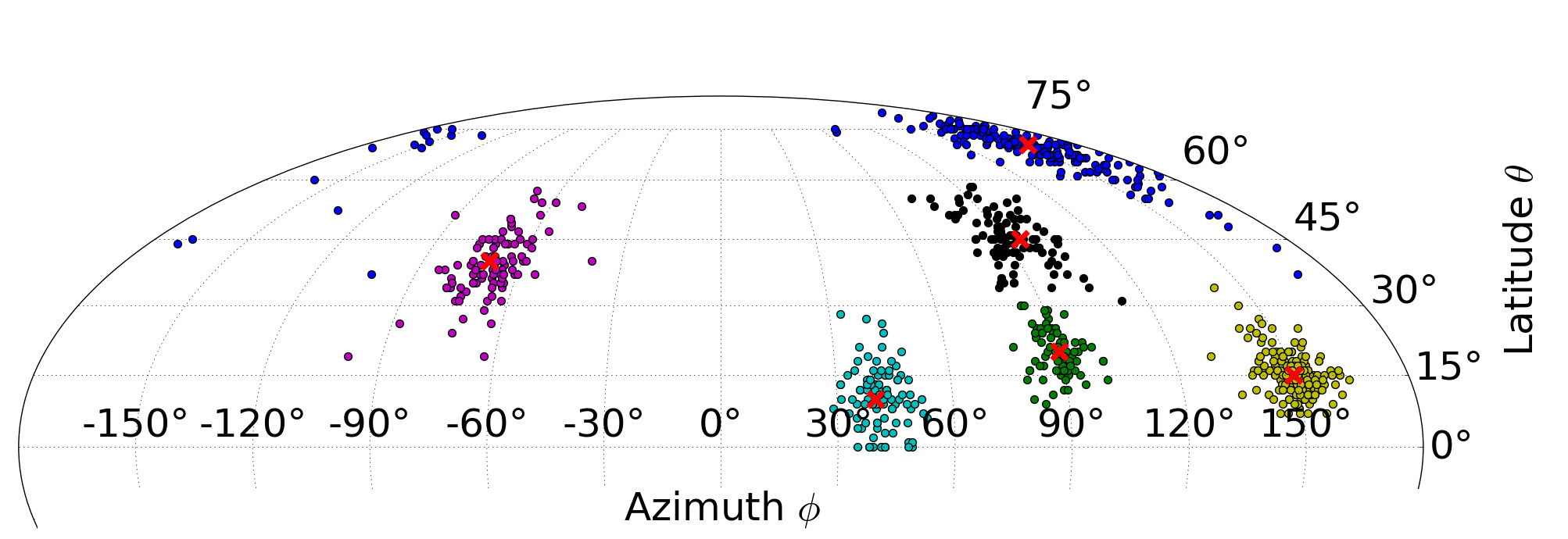}
\includegraphics[scale=0.25]{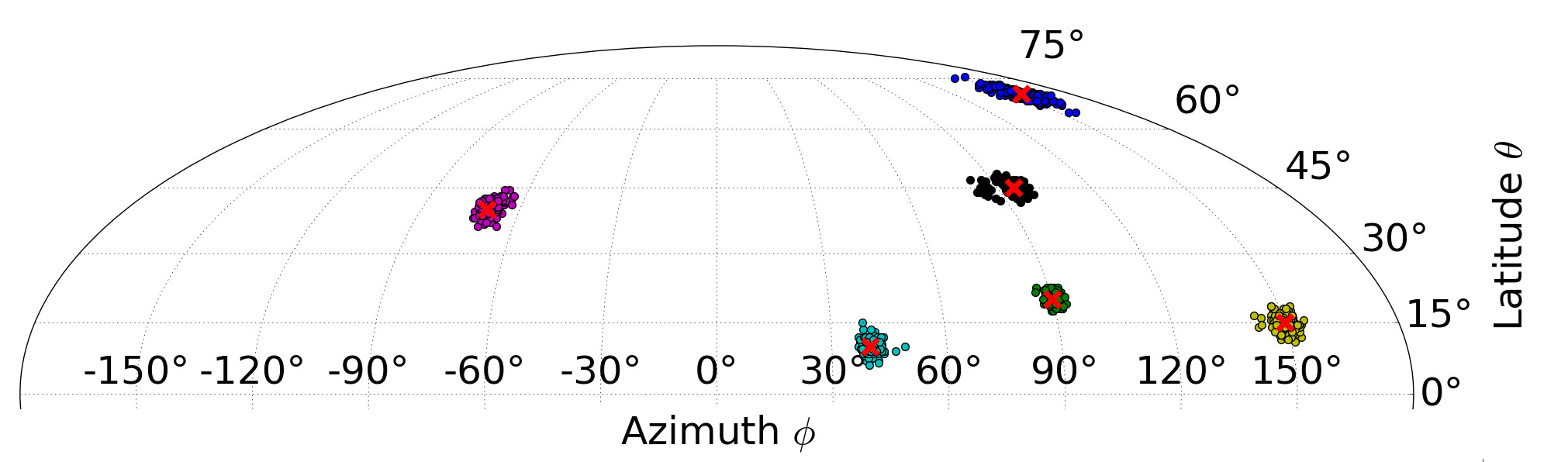}
\end{center}
\caption 
{ \label{fig:GRB_Sim} A comparison between the simulation-generated GRBs direction reconstructed by different detector configurations.  Each  dot represents  a  1 s burst of 10~ph cm$^{-2}$ s$^{-1}$ at lower earth orbit. Dots are grouped by color according to the actual  burst  direction,  which  is  represented by  a $\times$ mark.   \textbf{Top}: GTM. \textbf{Middle}: 90-scintillator GALI. \textbf{bottom}: 350-scintillator GALI. Notice the clear improvement in localization accuracy when the number of scintillators is increased.}
\end{figure}

A small sample of the generated bursts' direction reconstructed by the three systems is shown in figure \ref{fig:GRB_Sim}. The all-sky average deviation of the reconstructed source direction from the original source direction in each system is 13.1$^\circ$ for the GTM, 6.5$^\circ$ for a 90-scintillator GALI and 1.7$^\circ$ for the 350-scintillator GALI.
From these simulations we conclude that the localization accuracy of a system increases significantly with the number of detectors, even when the effective area of the entire system is reduced.

\section{Experimental setup}
\label{sec: laboratory experiments}

\subsection{Individual detector unit characterization}
\label{subsec: detector unit}
We conducted a preliminary study to characterize single detector units in order to choose among different scintillating crystals, reflective wrappings and SiPM sensors. Each detector unit consists of a (9\,mm)$^3$ cubic scintillator wrapped or coated by a reflective material and coupled to a \sipm~using Cargille Meltmount\footnote{\texttt{https://www.cargille.com/mounting-media}} optical glue, which has a refractive index of 1.7. 

\paragraph{Scintillating crystals} of two kinds were selected for their high density and light yield, as well as their negligible intrinsic radioactivity: \csi~(see for example Balamurugan et al.\cite{balamurugan2006growth}) and \gagg\cite{kamada20122}. Both scintillators were polished by and purchased from Advatek\footnote{\texttt{https://www.advatech-uk.co.uk}}. We coated the \csi~crystals by sputtering a 50~nm thick layer of \sio, which protects the delicate and slightly hygroscopic crystal from mechanical degradation and water absorption.
\gagg~suffers from afterglow for hours after exposure to light due to lattice defects\cite{yoneyama2018evaluation}; therefore, it should be always kept in the dark so as not to alter the measurements. For space applications, the afterglow caused by electron and $\gamma$-ray dose needs to be considered.

\begin{figure}[htbp]
\begin{center}
\subfloat{
\includegraphics[scale=0.4]{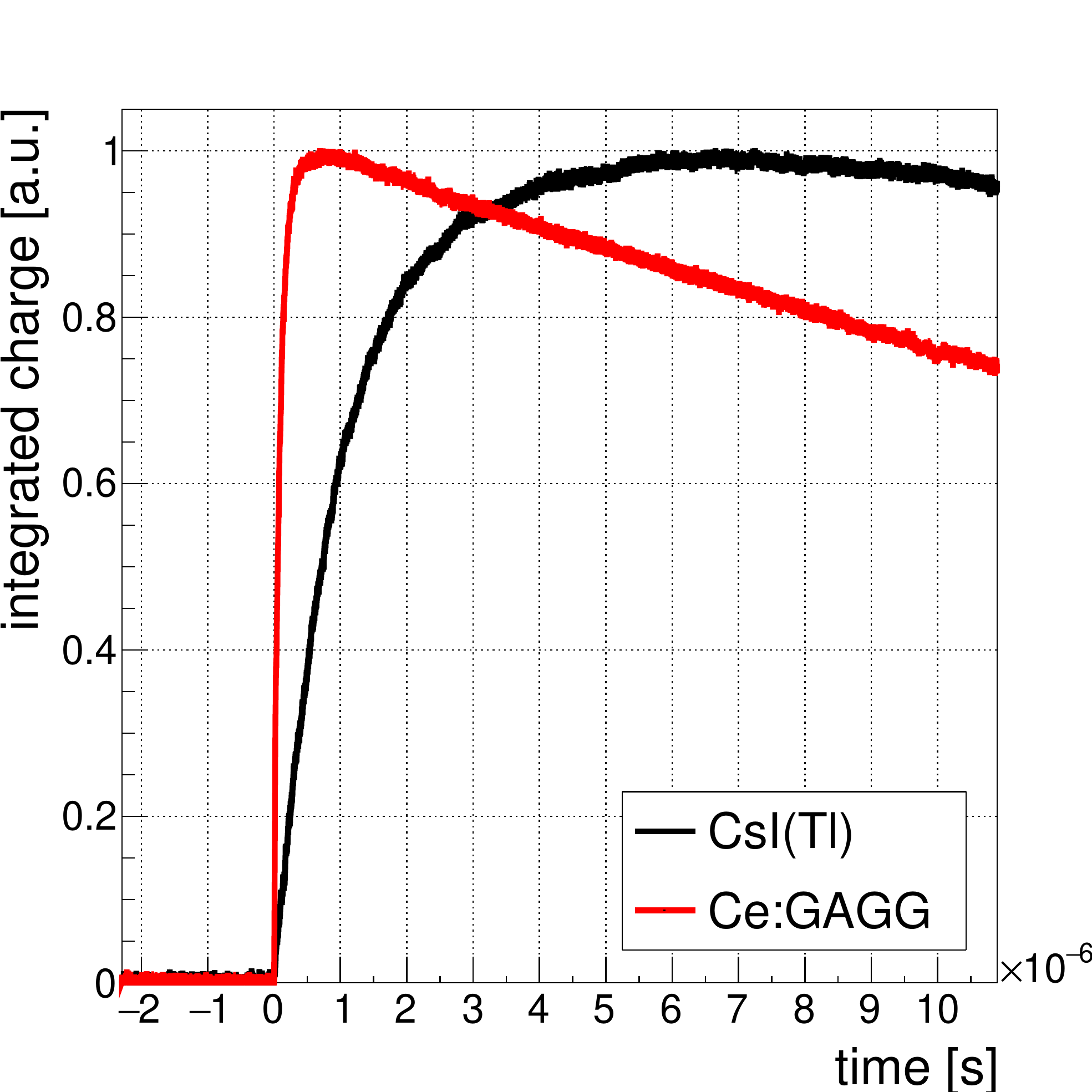}
\label{subfig: waveforms}
}
\subfloat{
\includegraphics[scale=0.4]{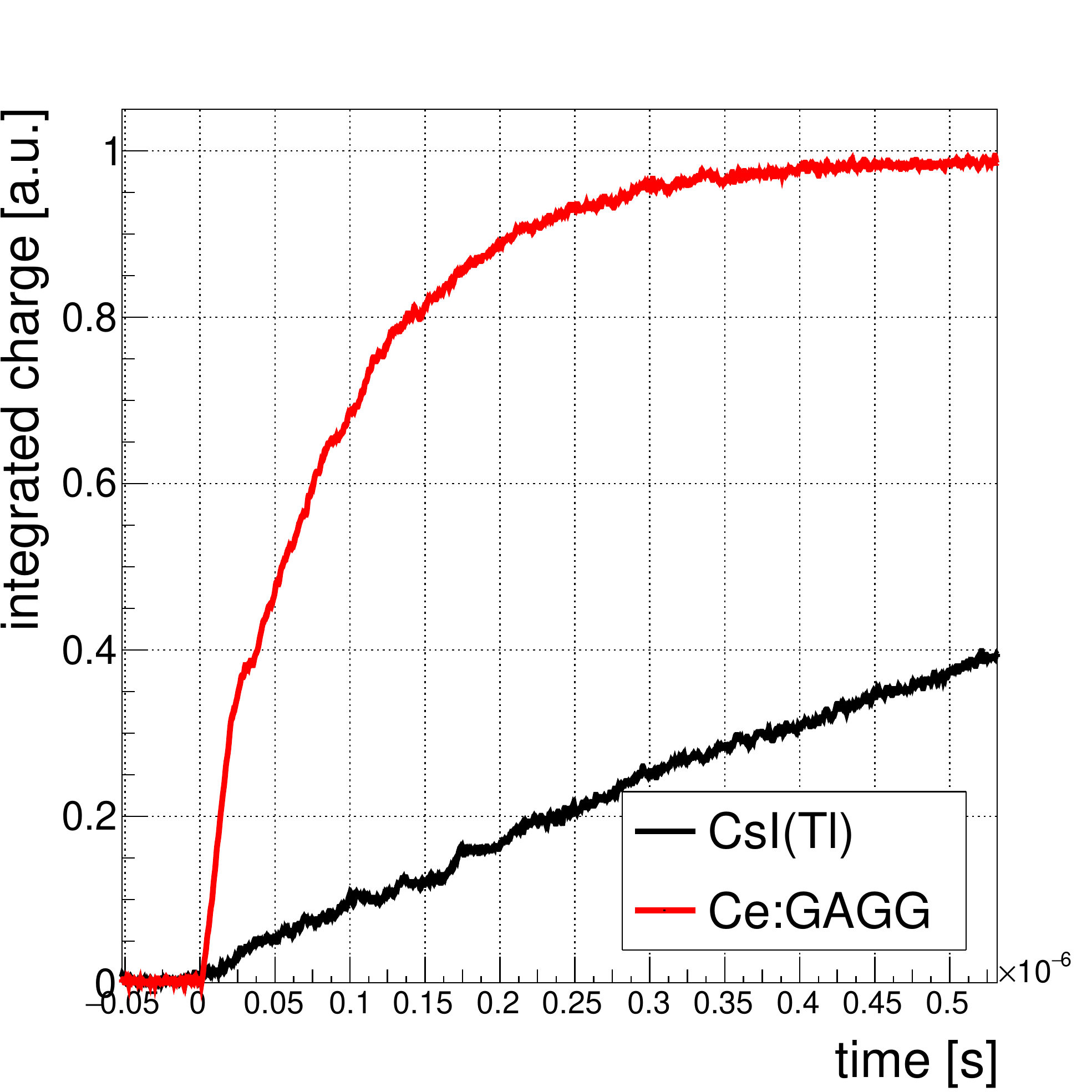}
\label{subfig: waveforms zoom}
}
\end{center}
\caption{$\gamma$-ray signals  from \csi~and \gagg~crystals read by a \sipm~coupled to ground through a 10~k$\Omega$ resistor. The right hand side shows the first 500~ns of the plot on the left. The measured voltage is proportional to the integrated charge produced by the \sipm. The peak corresponds to the end of the light production, which is faster for \gagg. The voltage recovery follows the circuit RC value.}
\label{fig: crystal signals}
\end{figure}

$\gamma$-ray signals from our \csi~and \gagg~crystals are read by a \sensl~SiPM\footnote{\texttt{https://www.onsemi.com/pub/Collateral/MICROJ-SERIES-D.PDF}} through a 10~k$\Omega$ resistor and recorded with an oscilloscope with 1~M$\Omega$ termination, and are shown in figure~\ref{fig: crystal signals}. In these measurements the \sipm~was operated at a 27.2~V bias. 
The recorded voltage pulse is a convolution of the \sipm~response function and the scintillation time evolution\cite{spanoudaki2011scintillation}. For both crystals in the present configuration (RC$\sim$40~$\upmu$s, given C$\sim$4~nF) the scintillation time is shorter than the circuit typical decay time. Therefore, the measured voltage is proportional to the integrated charge produced by the \sipm, and the peaks in figure~\ref{fig: crystal signals} corresponds to the end of the scintillation light production. The voltage recovery follows the circuit RC value.
Figure~\ref{subfig: waveforms zoom} shows the full \gagg~light emission signal, which presents a fast component ($\sim$100~ns) and a slow tail (ending at $\sim$500~ns) that can be distinguished by the signal slope change. 
In the same figure it can be seen that the \csi~light emission is much slower than the \gagg~one. During the 500~ns in which the \gagg~emission is completed, \csi~produces only $\sim$40$\%$ of the total charge. In figure~\ref{subfig: waveforms} the full \csi~signal is shown, reaching its maximum value within $\sim$6~$\upmu$s. Most existing integrated readout electronics have much shorter integration times, which may hinder the sensitivity of \csi~scintillators. Considering a readout electronics having $\sim$1~$\upmu$s integration time (like the TOFPET2 ASICs described in section~\ref{subsec: 90 units detector prototype}) there is a clear advantage in using \gagg~for better signal-to-noise.

\begin{figure}[htbp]
\begin{center}
\includegraphics[scale=0.5]{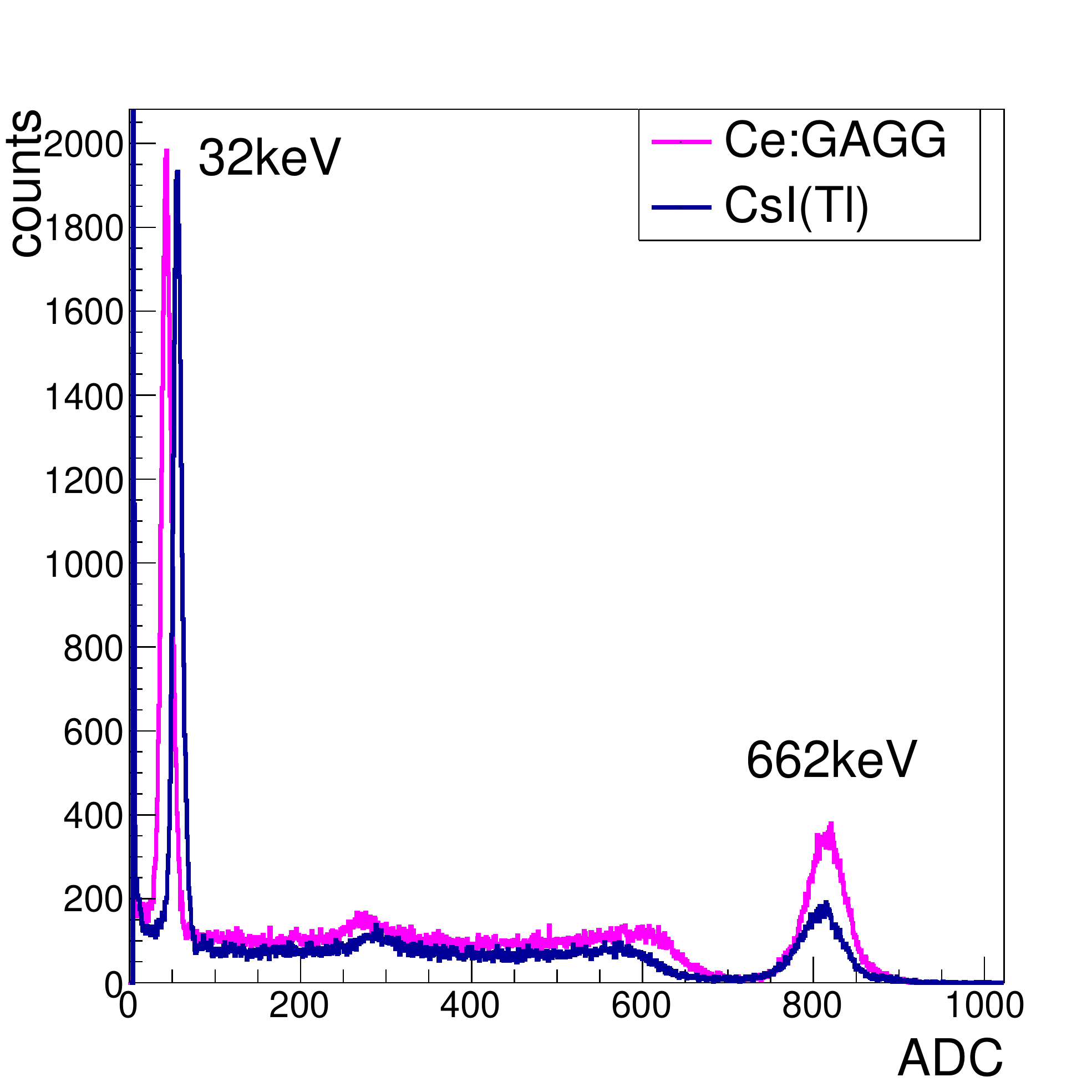}
\end{center}
\caption{Spectra from a \cs~source detected by (9\,mm)$^3$ \csi~and  \gagg~cubic scintillators. The 662~keV peaks clearly show the higher detection efficiency of \gagg~ at high energies.}
\label{fig: 137Cs spectra}
\end{figure}

We measured charge spectra produced by the two crystals using an Ortec 671 spectroscopy amplifier with an Ortec Aspec-927 MCA. We set the amplifier shaping time to 0.5~$\upmu$s for \gagg~and to 3~$\upmu$s for \csi~to include the full signal rise. The latter value is due to the signal shortening because of the low input  impedance (465~$\Omega$) of the amplifier. Spectra from a \cs~source from both scintillators are presented in figure~\ref{fig: 137Cs spectra}. The ratio of the number of counts in the 662~keV peak between the two scintillators is $\sim$2 in favor of \gagg, whereas for the 32~keV peak the ratio is comparable (taking into account the higher Compton baseline of \gagg). This indicates a higher detection efficiency of \gagg~due to its higher density. On the other hand, it has been reported\cite{sibczynski2017characterization} that at low energies \csi~provides $\sim$120\% response with respect to the response at high energy while \gagg~provides only $\sim$85\%. In conditions in which the signal to noise ratio is low, such as high temperature or high \sipm~dark count rates (for example due to radiation damage), the noise threshold rises, so the advantage due to higher \gagg~efficiency can be jeopardized. In general, the spectra peak positions result from a few factors: the crystal light yield, the \sipm~gain and the effect of the amplifier. 
Investigating these factors and their $\gamma$-ray energy dependence requires a dedicated study with multiple radioactive sources, which is beyond the scope of the present study.

Based on the presented results we conclude that there might be advantages in choosing \gagg~for its faster light emission and higher detection efficiency. Despite that, for our first system prototype we chose \csi~because of its lower cost. Next we plan to build a full \gagg~prototype to compare their performance.

\paragraph{The \sipm} models that we considered are: \sensl~and \hama\cite{yamamoto2019recent}\footnote{\texttt{\detokenize{https://www.hamamatsu.com/resources/pdf/ssd/s14160_s14161_series_kapd1064e.pdf}}}. The most relevant features for the two kinds are summarized in table~\ref{tab: sipm}. The \sensl~\sipm~ has two clear advantages: a 13.5~V lower breakdown voltage and a smaller cell size, which implies a linear response until higher gamma energies. On the other hand, \hama~has lower dark current, which is important for space applications. Additionally, according to vendor data the \hama~has a $\sim$10$\%$ enhanced PDE in the region above 450~nm, where the emission of both scintillators peaks.

\begin{center}
\begin{tabular}{ |c|c|c|c|c|c|} 
 \hline
 Type & size & $V_{bd}$[V] & max I$_{dark}$[$\upmu$A& cell size [$\upmu$m] & capacitance [nF]\\
 \hline
 \sensl & 6~mm & 24.5 & 12 & 35 & 4 \\
 \hline
 \hama & 6~mm & 38 & 7.5& 50 & 2  \\  
 \hline
\end{tabular}
\label{tab: sipm}
\end{center}



Another important parameter for space applications is radiation hardness, meaning capability to withstand doses of highly ionizing particles with an acceptable performance degradation. We could not find any study of performance degradation due to intense irradiation for \sensl, whereas recent studies were published for \hama~ within the context of the CAMELOT mission~\cite{vripa2019estimation}. One study shows that for \hama~the performance degradation - in terms of dark current and noise threshold - due to 200~MeV protons recovers over time, as it undergoes an annealing process at room temperature~\cite{hirade986annealing}. Another study reports that heavy-ion irradiation causes an increased dark current and a worse energy resolution, but the effect is overall mild~\cite{link2019silicon}.

Clearly, more testing of radiation damage on \sipm~is required. Nonetheless, taking all this into account we chose \hama~for our prototype.

\paragraph{Reflective materials} of various kinds can be applied to the crystal faces in order to maximize the light collection from the crystal to the \sipm. The reflectors considered here are: Teflon tape (diffusive), Ag-Al coating (specular), and 3M-Vikuiti$^{TM}$ enhanced specular reflector (ESR). 

\begin{figure}[htbp]
\begin{center}
\subfloat{
\includegraphics[scale=0.3]{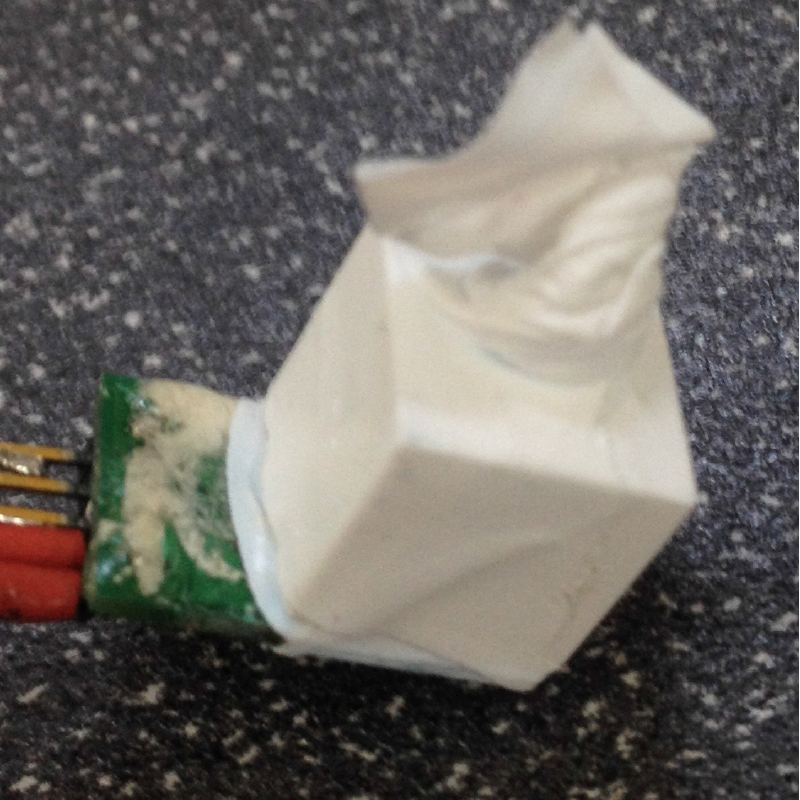}
\label{subfig: teflon}
}
\subfloat{
\includegraphics[scale=0.3]{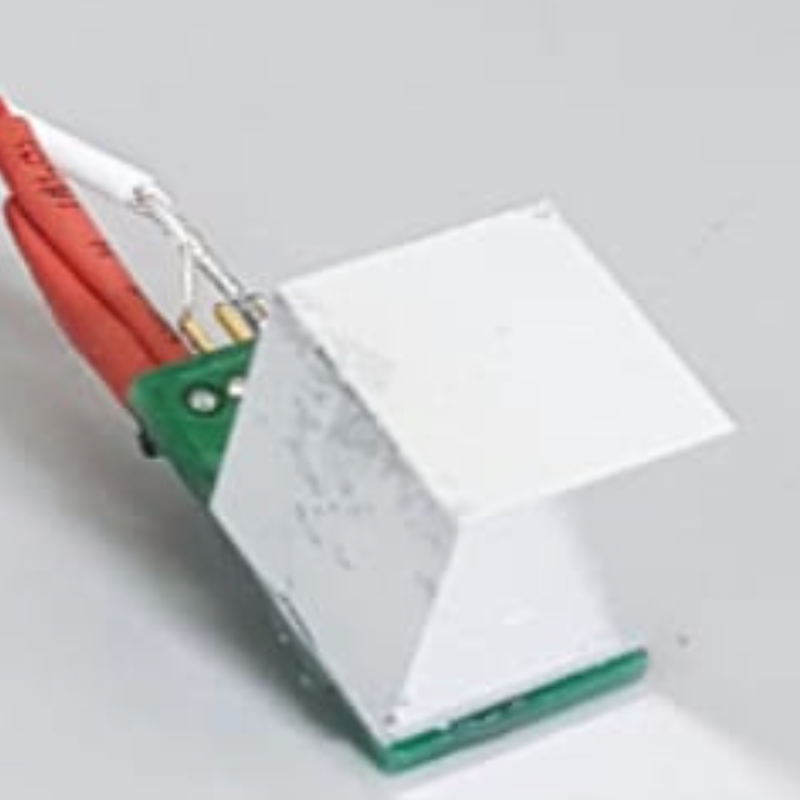}
\label{subfig: reflective coating}
}
\end{center}
\caption{\textbf{Left:} A \csi~crystal coupled to a \sipm~and then wrapped in Teflon tape. \textbf{Right:} A \csi~crystal coated with Al and Ag and coupled to a \sipm.}
\label{fig: reflectors}
\end{figure}

The Teflon tape is 80~$\upmu$m thick and needs to be winded a few times around the crystal, which is glued onto the \sipm, as shown in figure~\ref{subfig: teflon}. We discarded Teflon wrapping because of its large thickness and the difficulty of applying it uniformly and repeatably in the same fashion for each unit.

We sputtered \sio-coated \csi~crystals with a 200~nm Ag layer and then a more robust 200~nm Al layer on top of it. A 6~mm square area at the center of one cube face was left open to couple the \sipm. An example of such a unit is shown in figure~\ref{subfig: reflective coating}.

\begin{figure}[htbp]
\begin{center}
\subfloat{
\includegraphics[scale=0.3]{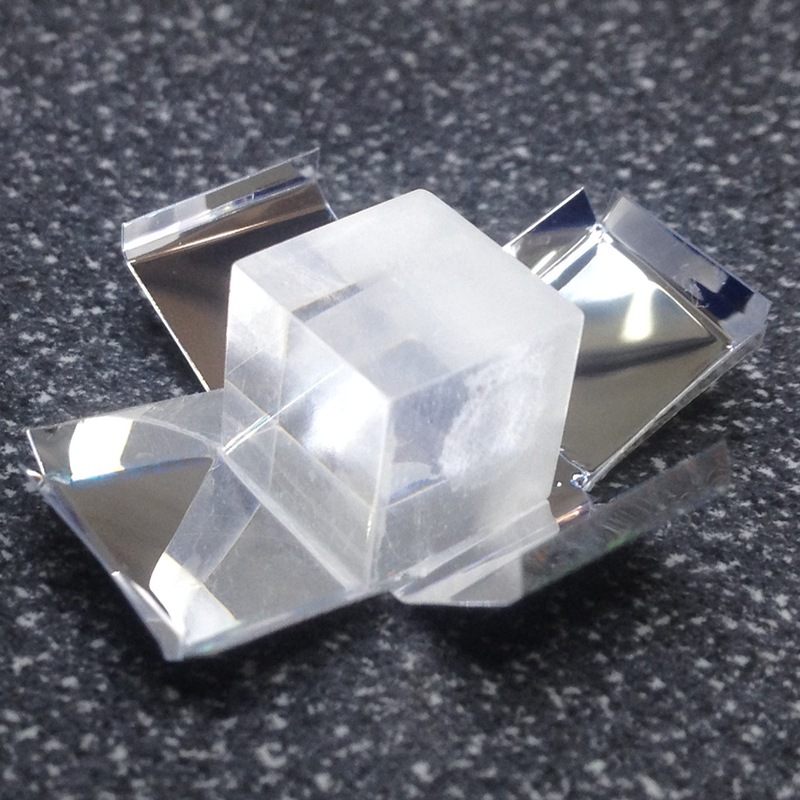}
}
\subfloat{
\includegraphics[scale=0.3]{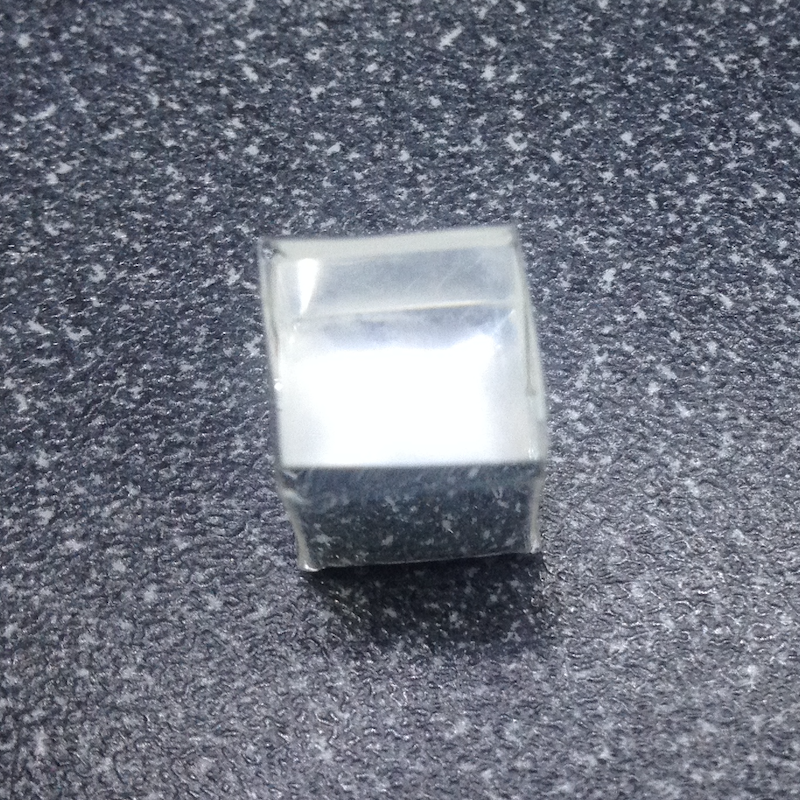}
}
\subfloat{
\includegraphics[scale=0.3]{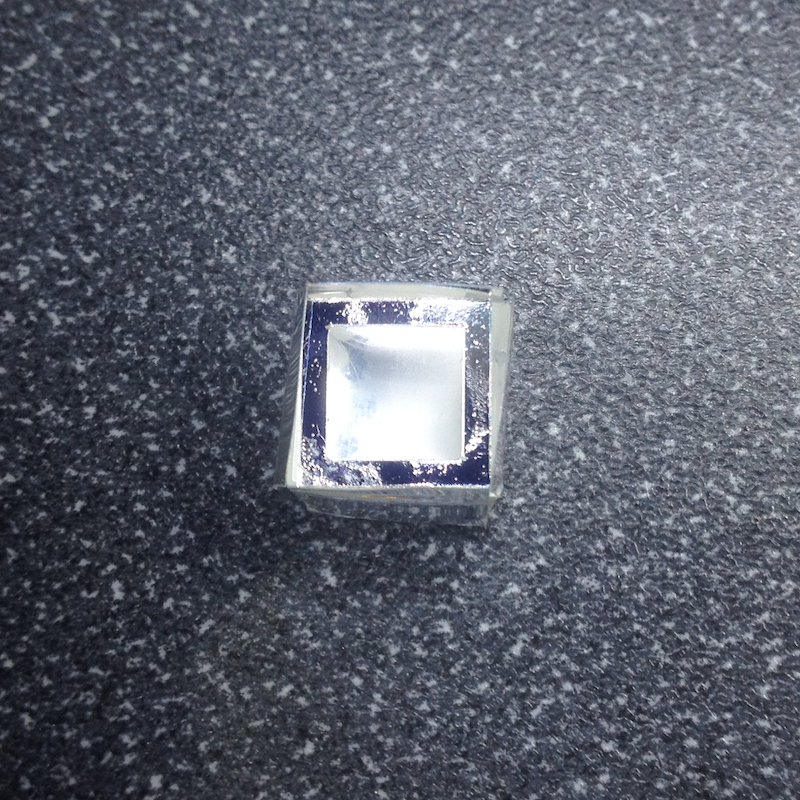}
}
\subfloat{
\includegraphics[scale=0.3]{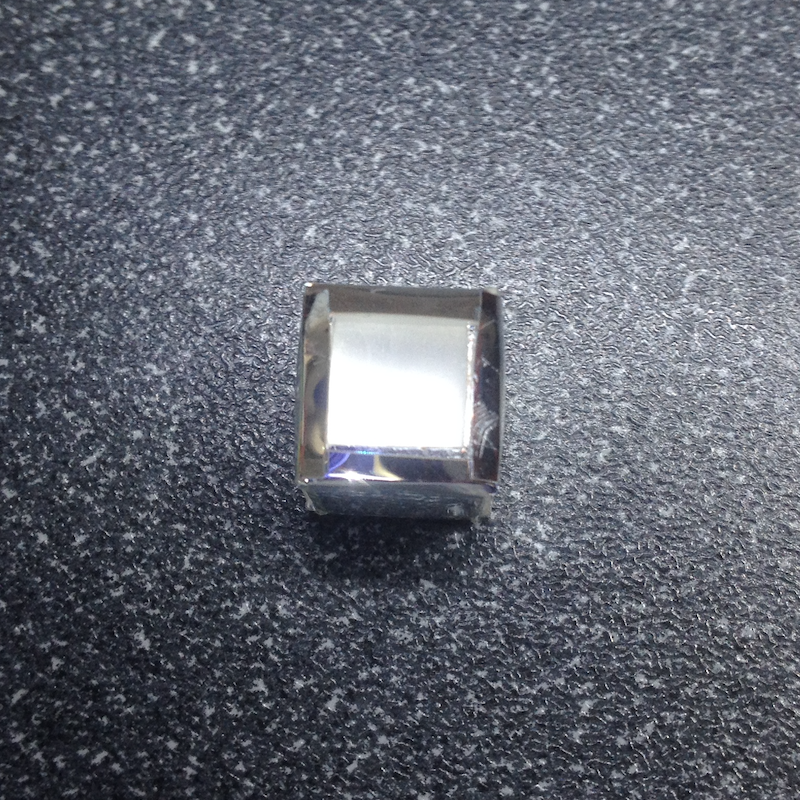}
}
\end{center}
\caption{From left to right, the \csi~crystal wrapping process with Vikuiti ESR.}
\label{fig: vikuiti}
\end{figure}

The Vikuiti ESR is a 65~$\upmu$m thick film. It was laser-cut and  glue\footnote{3M-9471} was applied to specific areas to keep it closed after folding. The wrapping process is illustrated in figure~\ref{fig: vikuiti}.

\begin{figure}[htbp]
\begin{center}
\includegraphics[scale=0.5]{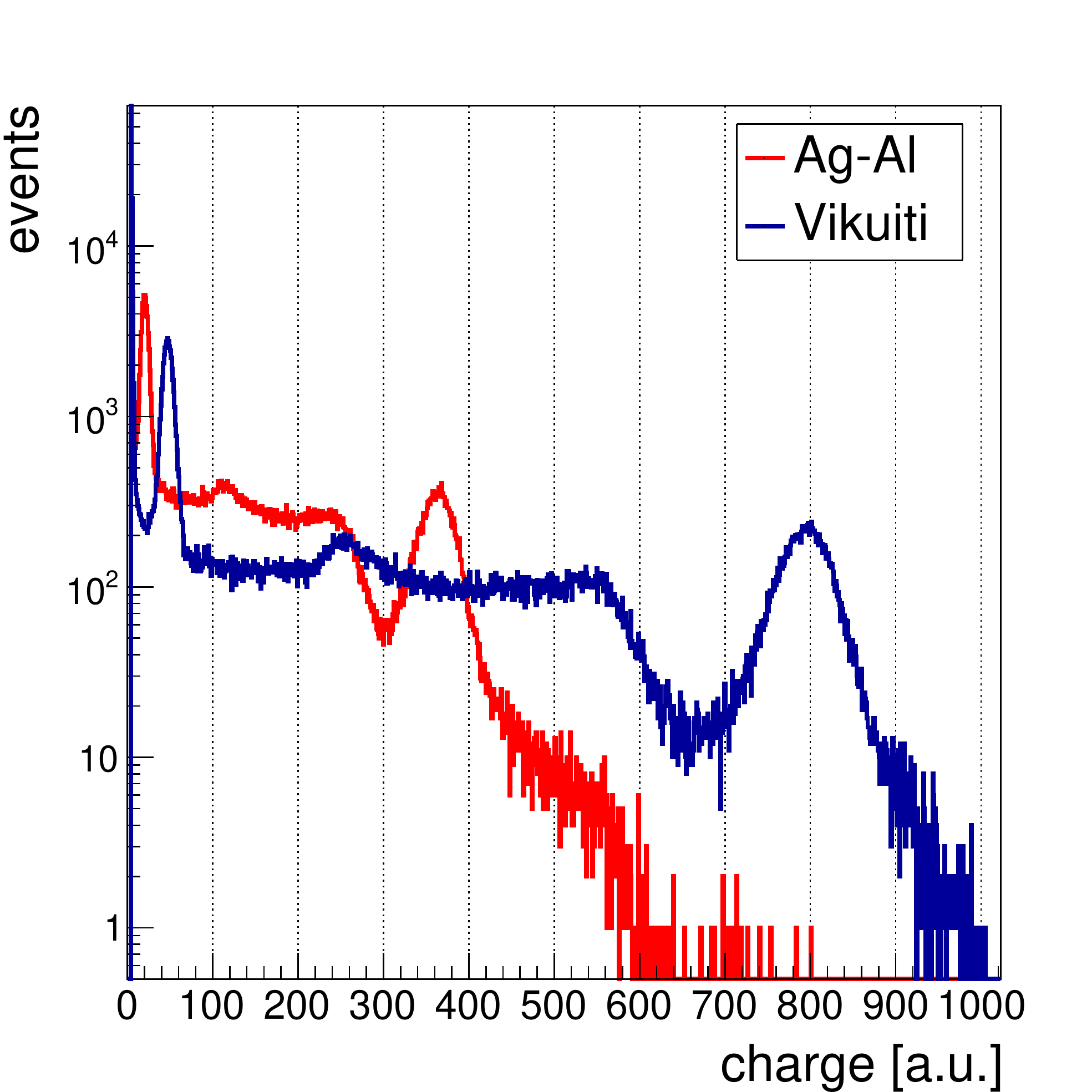}
\end{center}
\caption{Comparison of a \cs~spectrum measured by a \csi~crystal coated with Ag and then Al, and one measured by a Vikuiti ESR wrapped crystal. The light transport in the coated crystal results in noticeably less charge.}
\label{fig: reflectors spectra}
\end{figure}

A comparison of a \cs~spectrum measured by a \csi~crystal coated with Ag and Al with one measured by a Vikuiti ESR wrapped crystal is presented in figure~\ref{fig: reflectors spectra}. From the ratio of the 662\,keV peaks position, one can see that the light transport in the coated crystal results in a $\sim$2.2 times lower signal than the Vikuiti ESR wrapped one. This is surprising, given the fact that both Ag and Al have a reflectivity higher than 90$\%$ in the \csi~spectral range. This effect deserves further investigation. Since we repeated a similar experiment with a \gagg~crystal obtaining similar results, we conclude that the cause is probably not a chemical reaction that alters the properties of the surface. For Vikuiti ESR wrapping there is a layer of air between the crystal and the reflector which causes internal reflection to take place. This may affect the results, but it is not clear if this fact alone can explain the significant difference from the coated crystal.

\subsection{GALI detector prototype}
\label{subsec: 90 units detector prototype}
\begin{figure}[htbp]
\begin{center}
\subfloat{
\includegraphics[scale=0.4]{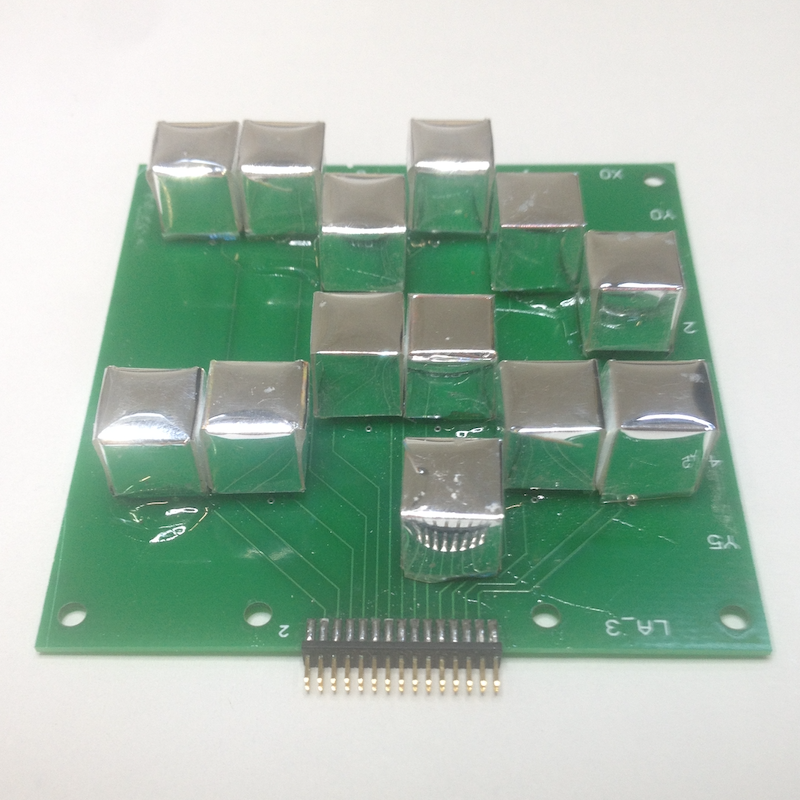}
\label{subfig: gali layer}
}
\subfloat{
\includegraphics[scale=0.202]{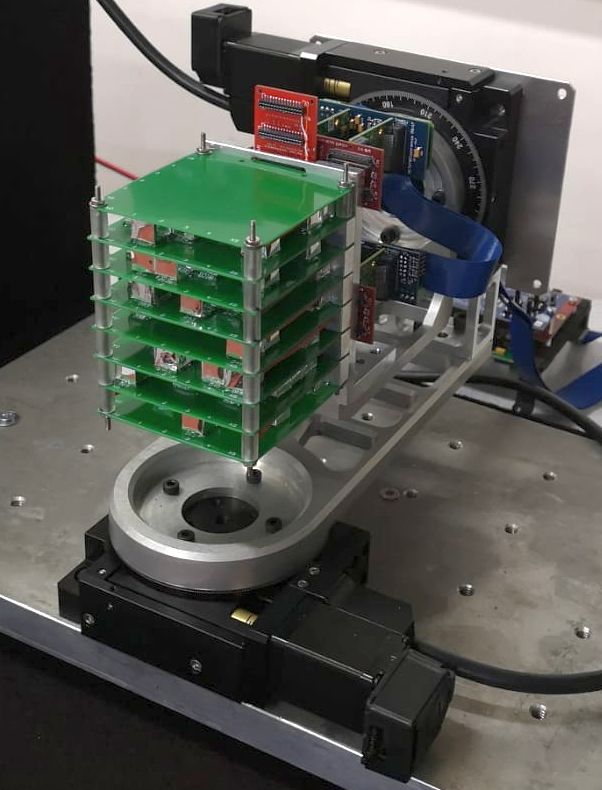}
\label{subfig: gali}
}
\end{center}
\caption{\textbf{Left:} A GALI detector layer composed of 13 units assembled on a PCB board. \textbf{Right:} The present prototype, equipped with PETSys DAQ system. The assembly is installed on motorized axes, which allows automatic angle scanning.}
\end{figure}

Our first GALI prototype is a system of 90 detector units assembled in the way described in section~\ref{subsec: detector unit}. Each \csi~crystal is coated with \sio, wrapped in Vikuiti ESR film and coupled to a \hama~\sipm~with optical glue. The detector units are arranged on 7 layers of standard FR4 PCB boards, each hosting 13 or 12 units. First, the \sipm s are soldered onto the PCBs, then the wrapped crystals are glued to them by means of a positioning jig. During the gluing stage the PCB and the glue dispenser are kept in an oven so that the glue melts. An assembled layer is shown in figure~\ref{subfig: gali layer}. The layers are then stacked by means of poly-carbonate rods and spacers and aluminium holders that allow mounting the entire stack on motorized axes for automatic angle scanning. On the back of the vertical axis a PETSys DAQ system is mounted\footnote{\texttt{https://www.petsyselectronics.com}}: a scalable readout based on the TOFPET2 ASICs. Each detector layer is connected to one of the two ASICs front-end boards through a PCB adapter. The front-end boards are connected to the DAQ through flat cables. 
Each ASIC has 64 channels independently connected to a single \sipm. On top of providing bias voltage to the \sipm, each channel integrates its output current and digitizes the integral whenever a threshold value is passed. The raw data are stored conveniently in a ROOT\footnote{\texttt{https://root.cern.ch}} tree for offline analysis. The data acquisition software can be controlled by python\footnote{\texttt{https://www.python.org}} scripts, as well as the rotating axes, so that the measurements are fully automated.

The entire assembly is shown in figure \ref{subfig: gali}. We installed the assembly in a thermally insulating dark box where back-fed peltier plates keep the inner temperature constant at 24$\pm 0.5^{\circ}$C. This is crucial for a stable gain of the \sipm s and of the DAQ.

\section{Localization measurements and results}
\label{sec:results}
In order to test the direction reconstruction capabilities of GALI we expose it to a 10 mCi \am~source placed approximately 3.5 meters away to simulate a distant source. The effective flux of the 59.6~keV line at this distance is approximately 50~ph cm$^{-2}$ s$^{-1}$. We scan the entire hemisphere by varying $\theta$ between 0 and 90$^{\circ}$ and $\phi$ between 0 and 360$^{\circ}$ with 5$^{\circ}$ intervals.  For each angle we acquire two kinds of measurements. First a 60\,s long exposure (corresponding to up to $\sim 3000$ counts on each scintillator), and then a series of 100 bursts 0.5\,s long. Measurements are analyzed offline. We use the long measurements to define the spectrum region of interest for each detector unit, namely, 2$\sigma$ region around the 59.6~keV Gaussian peak. The long measurements also provide the average counts in each detector at each angle. These are then interpolated at 0.5$^\circ$ intervals. We then test the GALI direction reconstruction accuracy of the bursts, similar to the method described for the simulations (section \ref{sec:sim}). 

To quantify the advantages of the new concept, we compare GALI to the traditional design used in the GTM. The results of the 90-scintillator GALI are compared to those obtained by a GTM prototype composed of four 6.35$\times$6.35$\times$2.54~cm$^3$ box NaI scintillators. For practical reasons the GTM experiment is limited to a quarter of the hemisphere. The GTM system is exposed to the source for 30\,s. Due to the larger effective area of each scintillator, a 30\,s exposure (corresponding to $\gtrsim 50000$ counts on each scintillator), is enough to establish the reference average counts. We Test the GTM localization capabilities using 50 bursts of $\sim 0.5\pm0.05$\,s. This uncertainty in burst length is caused by limitations of the DAQ system (CAEN DT5724\footnote{\texttt{https://www.caen.it}}). The GTM data are analyzed in a similar manner to the GALI data as described above.

\begin{figure} [!htbp]
\begin{center}
\includegraphics[scale=0.5]{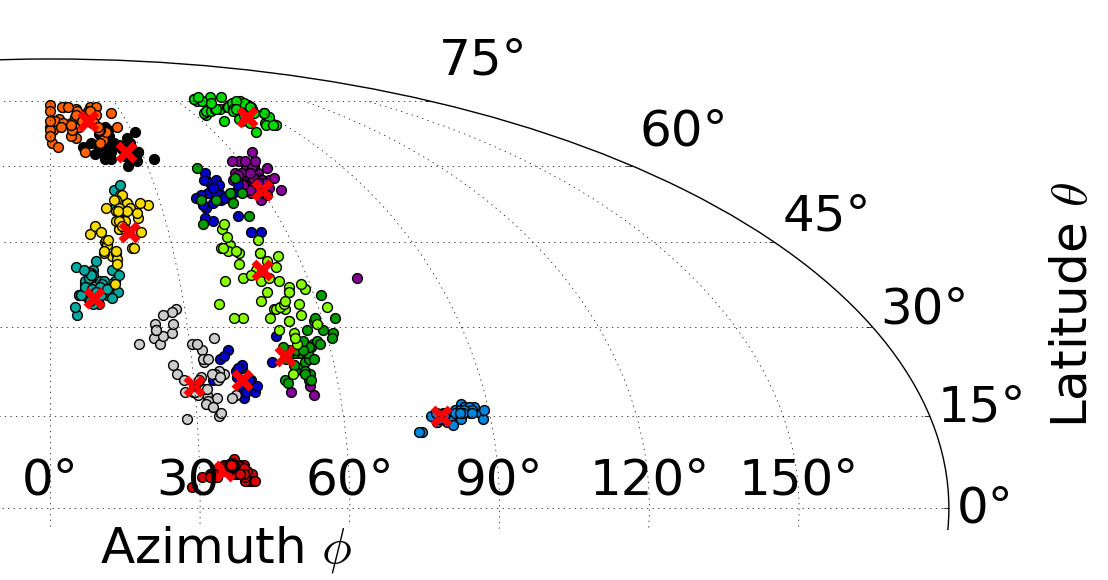}
\includegraphics[scale=0.5]{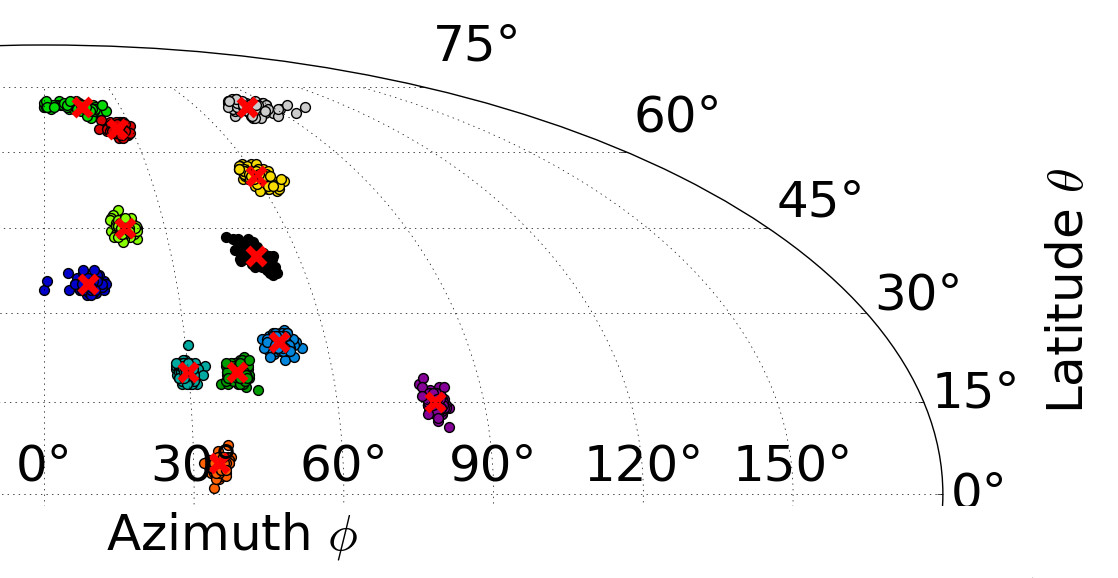}
\end{center}
\caption 
{ \label{fig:exp_res} A comparison between laboratory tests of the GTM  (\textbf{Top}) and a 90-scintillator GALI (\textbf{Bottom}). The reconstructed direction of repeated 0.5~s bursts are plotted where each dot represents a burst. Dots are grouped by color according to the actual source direction, which is represented by a $\times$ mark. Confirming the simulations, the superior angular localization accuracy of GALI is clear. }
\end{figure}

A comparison between the performance of the GTM system and the 90-scintillator GALI is shown in figure \ref{fig:exp_res} for 12 different burst directions. As can be seen, the 90-scintillator system demonstrates better accuracy overall. The average error for the shown reconstructed bursts ranges between 1.3$^\circ$ and 2.8$^\circ$ for the 90-scintillator GALI and between 3$^\circ$ and 21$^\circ$ for the GTM. GALI performs strictly better in all measured directions.  The improvement in accuracy is often greater than 50\%, despite the significantly smaller total detecting volume of GALI with respect to the GTM.

\section{Conclusions}
The present experiments can be summarized as follows

\begin{itemize}
\item 
We present a novel directional GRB  detector concept based on mutual occultation of numerous, small scintillator elements. 

\item
For this purpose, we explored both \csi\ and \gagg\ scintillators, as well as their coating and wrapping procedures. Two \sipm\ readout elements were also compared. We chose for the current experiment to use \csi\ coated with \sio\ and wrapped by a Vikuiti ESR reflector.

\item
We built a laboratory prototype consisting of 90 \csi\ (9\,mm)$^3$ cubes stacked in 7 layers. Relative count rates were measured at angles over the entire hemisphere with long (60\,s) exposures, and used as reference. Subsequently, the direction reconstruction capability was tested on short bursts (0.5\,s).

\item
Experimental results with the 59.6\,keV peak of \am~ show that short bursts ($\sim$25 ph\,cm$^{-2}$) can be localized to within 1$^\circ$--3$^\circ$. 

\item
Simulations of a 350-scintillator instrument, including true LEO background and GRB spectra, show an accuracy of $\sim1.7^\circ$, which outperforms any existing scintillator-based GRB instrument that we are aware of, including much bigger ones.

\end{itemize}

\acknowledgments 
We acknowledge the work of project students Ori Zaberchik and Joseph Mualem in the laboratory. We thank Guy Ankonina for proficiently coating the crystals.
We acknowledge support by a grant from ISA, a Center of Excellence of the ISF (grant No. 2752/19), and a grant from the Pazy Foundation. R.R. is supported by a Ramon scholarship from the Israeli Ministry of Science and Technology.
\bibliography{report} 
\bibliographystyle{spiebib} 

\end{document}